\pdfoutput=1
\documentclass{aa}
\usepackage{float,graphicx,textcomp}
\usepackage{natbib,amsmath,amssymb}
\usepackage{txfonts}
\usepackage[utf8]{inputenc}
\usepackage{color}
\usepackage{epsfig}
\usepackage{bbm}

\begin{document}

\title{Stellar dynamos with $\vec{\Omega}\times \vec{J}$ effect}
\titlerunning{Stellar dynamos with $\vec{\Omega}\times \vec{J}$ effect}

\author{V. V. Pipin \inst{1} \and N. Seehafer \inst{2}}

\institute{Institute for Solar-Terrestrial Physics,
Siberian Division of the Russian Academy of Sciences, 664033 Irkutsk,
Russia\\
\email{pip@iszf.irk.ru}
\and
Institut f\"ur Physik und Astronomie, Universit\"at Potsdam,
	     Karl-Liebknecht-Str. 24/25, 14476 Potsdam, Germany\\
             \email{seehafer@uni-potsdam.de}
 }

\date{Received .......... /Accepted ..........}

\abstract
{
The standard dynamo model for the solar and stellar magnetic fields is based on the
$\alpha\Omega$ mechanism, namely, an interplay between differential rotation
(the $\Omega$ effect) and a mean electromotive force generated by helical turbulent
convection flows (the $\alpha$ effect). There are, however, a number of problems
with the $\alpha$ effect and $\alpha\Omega$ dynamo models. Two of them are that, in the case of the Sun,
the obtained cycle periods are too short and  the magnetic activity is not sufficiently
concentrated at low latitudes.
}
{We explore the role of turbulent induction effects that may appear in addition
to the $\alpha$ effect. The additional effects result from the combined action of rotation and an
inhomogeneity of the large-scale magnetic field. The best known of them is the $\vec{\Omega}\times\vec{J}$ effect. We also include anisotropic diffusion and a new
dynamo term which is of third order in the rotation vector $\vec{\Omega}$.
}
{We study axisymmetric mean-field dynamo models containing differential rotation,
the $\alpha$ effect and the additional turbulent induction effects.
The model calculations are carried out using the rotation profile of the Sun as
obtained from helioseismic measurements and radial profiles of other quantities
according to a standard model of the solar interior. In addition, we consider a dynamo
model for a full sphere which is solely based on the joint induction effects of rotation and an inhomogeneity of the large-scale magnetic field,
without differential rotation and the $\alpha$ effect (a $\delta^{2}$ dynamo model). This kind of dynamo model may be relevant for fully convective stars.
}
{
With respect to the solar dynamo, the inclusion
of the additional turbulent induction effects increases the period of
the dynamo and brings the large-scale toroidal field closer
to the equator, thus improving the agreement of the models with the observations.
For the $\delta^2$ dynamo working in a full sphere, we find dynamo modes which
are steady if the effect of anisotropic diffusion is not included.
The inclusion of anisotropic diffusion
yields a magnetic field oscillating with a period of the order of the turbulent
magnetic diffusion time.
}
{}

  \keywords{Stars: magnetic fields
	-- Sun: magnetic fields
	-- magnetohydrodynamics (MHD)               }

\maketitle

\section{Introduction}

Most solar and stellar dynamo models use the scenario proposed
by \citet{par55,par79} where the magnetic field
is produced by an interplay between differential rotation (the $\Omega$ effect) and the
collective action of turbulent cyclonic convection flows, widely known as the $\alpha$ effect \citep{stk66,krarad80}.
The scheme suggests that the $\alpha$ effect is responsible for the generation
of the poloidal component of the large-scale magnetic field (LSMF) of
stars and other cosmic bodies.

There is an ongoing debate on a number of problems connected with the $\alpha$ effect
and $\alpha\Omega$ dynamos
\citep[see, e.g.,][]{oss03,rudhol04,brasub05}.
For instance, 
the period of the solar activity cycle poses a problem. Namely, for mixing-length
estimates of the turbulent magnetic diffusivity in the convection zone and dynamo action
distributed over the whole convection zone, the
 obtained cycle periods are generally
much shorter than the observed 22\,yr period  of the activity cycle. 
For thin-layer dynamos the situation becomes even worse.

Furthermore, according to standard theory the strength of the $\alpha$ effect follows roughly a
$\cos\theta$ colatitude profile. That is, the effect is strongest near the poles.
Similarly, also the variation of the solar rotation rate with radius, responsible for the
$\Omega$ effect, is strongest at high latitudes
\citep{schouetal98}. But solar
magnetic activity in the form of active regions is mainly observed in latitudinal belts
relatively close to the equator. As a possibility to bypass this discrepancy,
meridional (poloidal) flows are
under discussion, leading to so-called flux-transport dynamos 
\citep[see, e.g.,][]{oss03,rudhol04,dikgil07}. Such flows may transport
toroidal magnetic flux toward the equator
and their speed may determine the cycle period.

In this paper, we consider the possible role of a turbulent dynamo mechnisms that may complement the $\alpha$ effect or may be an alternative to it. Namely,
according to mean-field dynamo theory, there are other turbulent sources of the LSMF besides
the $\alpha$ effect 
in  rotating electrically conducting fluids.
The influence of the turbulence on
the LSMF is expressed by the
mean turbulent electromotive force (MEMF) $\vec{\mathcal{E}}=\left\langle \vec{u}\times\vec{b}\right\rangle $,
where
$\vec{u}$ and $\vec{b}$ are the fluctuating parts of the velocity and magnetic
field (angular brackets denote averages).
Here we investigate the role of the $\vec{\Omega}\times\vec{J}$ effect \citep{rad69}
in axisymmetric mean-field
dynamo models  ($\vec{\Omega}$ is the angular velocity of the stellar rotation and $\vec{J}$ the large-scale
(or mean) electric-current density). This effect, which may be interpreted as resulting
from an anisotropic turbulent
electrical
conductivity, has been little investigated in the context of solar
and stellar dynamos.
In the commonly used representation of the MEMF
on the basis of symmetry arguments
 \citep[see][]{rad80,krarad80,rad00,radklerog03}, the $\vec{\Omega}\times\vec{J}$ effect
represents a contribution to the $\vec{\delta}$ term;
another contribution to this term is the shear-current effect \citep{rogkle03,rogkle04}.

An illustration of the $\vec{\Omega}\times\vec{J}$ effect
is given in Fig.~\ref{fig1}:
\begin{figure*}
\centering
\includegraphics[width=0.8\linewidth]{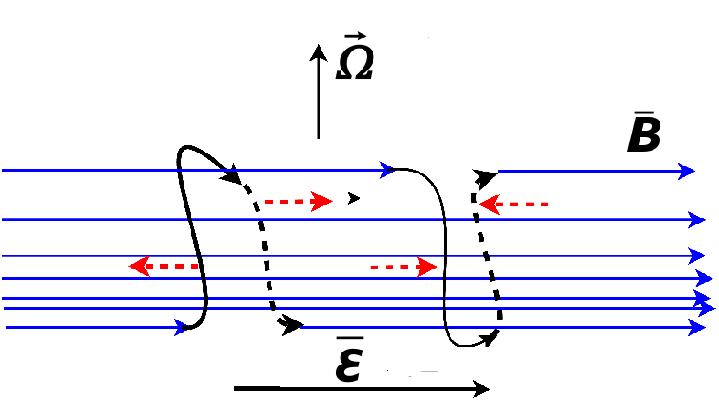}
\caption{\label{fig1}Schematic of the $\vec{\Omega}\times\vec{J}$
effect. }
\end{figure*}
The vector of the global rotation
points in the vertical direction, and the horizontal field lines of a purely toroidal LSMF
(solid blue lines) are linked by loop-like field lines (solid and dashed black lines) of a
magnetic perturbation. The strength of the toroidal LSMF (represented by the field line
density) decreases in the direction of the rotation vector.  The $\vec{\Omega}\times\vec{J}$ effect comes about via
the following steps. First, there is a contribution $(\vec{b}\cdot\nabla)\langle\vec{B}\rangle$ to the small-scale Lorentz force, giving rise to a velocity perturbation
$\vec{u}\sim\left(\vec{b}\vec{\cdot\nabla}\right)\langle\vec{B}\rangle$
parallel to the toroidal LSMF, indicated by
dashed red arrows. 
Second, the Coriolis force deflects the velocity perturbation in
the direction perpendicular to $\vec{\Omega}$ and $\langle\vec{B}\rangle$, that is, in the direction
perpendicular the plane of the figure in the schematic (in the radial and latitudinal
directions on the Sun or star). Third, the resulting deflected small-scale
flow and the small-scale magnetic field give rise to a
mean electromotive force $\vec{\mathcal{E}}\sim
\left\langle \left[\left(\vec{b}\vec{\cdot\nabla}\right)\left\langle \vec{B}\right\rangle \times\vec{\Omega}\right]\times\vec{b}\right\rangle
 \sim -\left\langle \vec{b}^{2}\right\rangle \left(\vec{\Omega\cdot\nabla}\right)\left\langle \vec{B}\right\rangle $, where in the last step we have assumed that $\vec{b}$ is mainly parallel
to $\vec{\Omega}$, so that a component of $\vec{\mathcal{E}}$ parallel to
$\vec{\Omega}$ could be neglected.
The induced electromotive force is then parallel to the LSMF.

The MEMF
was calculated analytically within a simplified version
of the $\tau$ approximation \citep[cf.][]{brasub05} by \cite{pip08}.
The relevant part
of the MEMF, containing the combined
effects of rotation and a nonuniformity of the LSMF,
reads
\begin{equation}
 \begin{split}
\mathcal{E}_{i}^{(d)}  ={}& \left\{ f_{1}^{(d)}e_{n}\overline{B}_{ni}+f_{2}^{(d)}\varepsilon_{inm}\overline{B}_{mn}+\varepsilon f_{3}^{(d)}e_{i}e_{n}e_{m}\overline{B}_{mn}\right. \\
 &+
 \left.f_{1}^{(a)}\varepsilon_{inm}e_{n}e_{l}
\left(2\varepsilon\overline{B}_{lm}-\left(\varepsilon+1\right)\overline{B}_{ml}\right)\right.
\\
&+ \left.\varepsilon f_{4}^{(d)}e_{n}\overline{B}_{in}\right\} \left\langle {\vec{u}^{(0)}}^2\right\rangle \tau_{c} \,,\label{dif1}
\end{split}
\end{equation}
where $\vec{e}=\vec{\Omega}/|\vec{\Omega}|$ is the unit vector in the direction of the
rotation vector,
$\overline{B}_{ij}=\partial\langle B_i\rangle/\partial x_j$ the gradient tensor
of the mean magnetic field
(we here use Cartesian coordinates $x_1,x_2,x_3$ and the summation convention),
$\vec{u}^{(0)}$ the small-scale or turbulent convective background velocity as present
in the absence of rotation and a mean magnetic field,
$\tau_{c}$ the correlation time of $\vec{u}^{(0)}$,
$\varepsilon=\sqrt{\left\langle {\vec{b}^{(0)}}^{2}\right\rangle }/\left(u_{c}\sqrt{\mu_0\rho}\right)$
the square root of the ratio between the energies
of a fluctuating magnetic background field $\vec{b}^{(0)}$, assumed to be generated by a
small-scale dynamo, and the background velocity field $\vec{u}^{(0)}$
($u_c=\sqrt{\left\langle{\vec{u}^{(0)}}^2\right\rangle}$ is the rms value
of the latter one and $\rho$ the mass density), and
$f_i^{(d)}$ and $f_i^{(a)}$ ($i=1,2,\dots$) denote functions of $\varepsilon$ and the Coriolis number
$\Omega^{\ast}$ that are given in the Appendix. $\Omega^{\ast}$,
measuring the influence of rotation on the turbulence,
 is defined by
$\Omega^{\ast}=2\Omega_0\tau_{c}$, with $\Omega_0$ denoting the
solid-body rotation rate; in our numerical calculations for the Sun we have chosen the value of the equatorial angular velocity at the solar surface for $\Omega_0$. Differential rotation,
which gives rise to the shear-current effect, is not included at this place.
We assume energy equipartition between the two background fields, i.e., $\varepsilon=1$.

By its nature, the assumed small-scale dynamo is fully independent of the mean field. An underlying assumption here is that developed turbulence in an electrically conducting, non-rotating fluid will be magnetohydrodynamic, rather than purely kinetic, if the small-scale motions are sufficiently complex. With rotation added, a mean-field dynamo can be superposed to the small-scale dynamo.

Hereafter, for simplicity, by the $\vec{\Omega}\times\vec{J}$ effect we mean all induction
effects in Eq.~(\ref{dif1}) which are of odd order in $\vec{e}$, that is, the terms with coefficients $f_1^{(d)}$,
$f_3^{(d)}$ and $f_4^{(d)}$, respectively.
Of the remaining two terms, with coefficients  $f_2^{(d)}$ and $f_1^{(a)}$, respectively,
that with coefficient $f_2^{(d)}$
corresponds to the $\beta$ term in commonly used representations of $\vec{\mathcal{E}}$ and describes
an isotropic turbulent diffusion. The action of the term  with coefficient $f_1^{(a)}$ will be
referred to as anisotropic diffusion. The second part of this term, proportional to $f_1^{(a)}(\varepsilon+1$), is known to
describe an extra diffusion along $\vec{e}$ \citep{kitpiprud94,kit02,kit04}.

The $\vec{\Omega}\times\vec{J}$ effect as it is usually understood is contained in the two source terms
of first order in $\vec{e}$, with coefficients $f_1^{(d)}$ and
$f_4^{(d)}$, respectively. Namely, in vector form the sum of these two terms can also be written as
\begin{equation}\begin{split} \label{first-order-terms}
&f_1^{(d)}\nabla(\vec{e}\cdot\vec\langle\vec{B}\rangle)+
\varepsilon f_4^{(d)}(\vec{e}\cdot\nabla)\langle\vec{B}\rangle \\
&=f_1^{(d)}\nabla(\vec{e}\cdot\vec\langle\vec{B}\rangle)+\varepsilon f_4^{(d)}[\nabla(\vec{e}\cdot\vec\langle\vec{B}\rangle)-\vec{e}\times(\nabla\times\langle\vec{B}\rangle)]\\
&=-\varepsilon f_4^{(d)}\vec{e}\times\mu_0\vec{J} +\left(f_1^{(d)}+\varepsilon f_4^{(d)}\right)
\nabla(\vec{e}\cdot\vec\langle\vec{B}\rangle)\,.
\end{split}\end{equation}
By their form, the two terms on the right-hand side of the last of Eqs.~(\ref{first-order-terms})
correspond to the $\delta_1$ (here, $\vec{\Omega}\times\vec{J}$) and $\delta_2$ effects,
respectively \citep[cf.][]{krarad80}.
In the axisymmetric case considered here, where gradients in the azimuthal direction
vanish, the $\delta_2$ effect does not contribute to the important azimuthal
component of the MEMF. Furthermore, it becomes a gradient, and its mean-field induction effect thus vanishes completely, if the Coriolis number $\Omega^*$ (i.e., the coefficient of
$\nabla(\vec{e}\cdot\vec\langle\vec{B}\rangle)$)
does not vary spatially
(as in a model considered in Sect.~\ref{sub:d2} below).

The term of third order in $\vec{e}$ in Eq.~(\ref{dif1}), with coefficient $f_3^{(d)}$, is an
additional source term that does not seem to have been included in dynamo studies before.
It involves only the symmetric part of the gradient tensor of the mean magnetic field
(since $e_ne_m\overline{B}_{mn}=e_ne_m\left(\overline{B}_{mn}+\overline{B}_{nm}\right)/2$)
and the coefficient of this symmetric part, $\varepsilon f_3^{(d)}e_ie_ne_m$, is symmetric in the
indices $m,n$.
By its formal structure it thus belongs to the $\kappa$ term in respresentations of the MEMF as used in more recent papers of R\"adler and his collaborators
\citep[see, e.g.,][]{rad00,radklerog03,radste06}.

First results on the $\vec{\Omega}\times\vec{J}$ effect in mean-field
dynamo models were, for instance, given in \citet{rad69}, \citet{sti76} and \citet{krarad80}. At that time the dependence of the
solar rotation rate on radius and latitude was not known
well enough. In the early papers also the possibility of combining the $\alpha\Omega$ and
$\delta\Omega$ mechanisms was discussed. Recently,
the idea of a combination of the $\alpha$ and $\vec{\Omega}\times\vec{J}$
effects in dynamo models was anew suggested by \citet{kit04}.
Here we explore axisymmetric kinematic $\alpha^2\delta\Omega$ dynamo models ($\alpha$ effect plus $\vec{\Omega}\times\vec{J}$ effect
plus differential rotation) for a convective spherical shell and for a full sphere, following the suggestion of \citet{kit04} and using the calculations of \citet{pip08}.

In the context
of the model the following points concerning the $\vec{\Omega}\times\vec{J}$
effect are important: 1) The effect generates a MEMF along the LSMF,
similar to the $\alpha$ effect. 2) The strength
of the $\vec{\Omega}\times\vec{J}$ effect depends both on the intensity of the fluctuating
magnetic fields and on their linkage with the LSMF. The latter contributes
to the amount of  magnetic and  current helicities in the volume considered.
Thus, the strength of the $\vec{\Omega}\times\vec{J}$ effect, as well as that of the $\alpha$
effect, is connected with the evolution of these  quantities. In a companion study
\citep{pip07}, the nonlinear saturation
of the $\alpha$ and $\vec{\Omega}\times\vec{J}$ effects was investigated by integrating the
mean-field equations coupled to an evolution equation for the small-scale current helicity
forward in time. Similar to the nonlinear back-reaction of the mean magnetic field on the $\alpha$ effect, known as $\alpha$ quenching, the $\vec{\Omega}\times\vec{J}$ effect may be suppressed by strong mean fields. Here we concentrate on the linear-stability problem. Rather than numerically
simulating time evolutions, the eigenvalues of the Jacobian matrix of the equations for the
mean magnetic field are calculated directly. For a comparison of model calculations for the $\alpha$ effect and helicities in the solar convection zone, carried out using similar approximations as for the calculation
of $\vec{\mathcal{E}}$ in the repesent study, with observations at atmospheric levels,
we refer to \citet{kuzpipsee06}. 

We construct an $\alpha^2\delta\Omega$ dynamo model for the Sun with distributed dynamo action in the
bulk of the convection zone. Such distributed dynamos have to be distinguished from
boundary-layer dynamos that operate in the overshoot layer at the bottom of the convection
zone, a thin transition region between the convection zone and the convectively stable radiative
core, which is believed to coincide with the tachocline, where
the differential rotation changes into rigid rotation in the radiative core.
The main argument in favour of boundary-layer dynamos is that in the convection zone proper,
due to the action of magnetic bouyancy, magnetic flux might not be stored long enough to allow
the generation of a sufficiently strong toroidal field by differential rotation. This argument
is based on the picture that the magnetic flux is concentrated in thin flux tubes
\citep[see, e.g.,][]{spiwei80,schus80,galwei81,schusfer03}. Large-scale magnetic fields
are only little affected by magnetic bouyancy \citep{kitpip93}. A critical discussion of arguments for and against deep-seated and distributed dynamos, respectively, is found in \citet{bra05}.

In addition, we develop a kinematic
model of an axisymmetric $\delta^{2}$ dynamo in a full sphere. This dynamo
is solely based on the turbulent electromotive force $\vec{\mathcal{E}}^{(d)}$ given by Eq.~(\ref{dif1}). It might be working in fully convective stars. These objects presumably do not
possess layers with a strong
velocity shear like the tachoclines at the bottom of the convection zones of solar-type stars, where
the toroidal part of the large-scale magnetic field is believed to be generated.
Nevertheless, fully convective stars are scarcely less magnetically active then stars with a radiative
core and should, thus, harbor dynamo action. Discussions of the resulting dynamo problem
may be found in the recent studies of \citet{dobstibra06}, \citet{chakuk06}
and \citet{bro08}.
Mean-field dynamo models for fully convective stars
have been mainly based on the $\alpha^2$ mechanism \citep{kukrud99,elsrud07,chakuk06}.
Also here, we suggest the $\vec{\Omega}\times\vec{J}$ effect (in its extended form) as a
complement or an alternative to the $\alpha$ effect.

The remainder of the paper is organized as follows: In Sect.~\ref{models} we describe our models
of the solar $\alpha^2\delta\Omega$ dynamo and the $\delta^{2}$ dynamo,
as well as the used numerical procedure. Then, in Sect.~\ref{results}, we present the obtained results.
Finally, in Sect.~\ref{sec_conclusions}, we draw conclusions and discuss our results.

\section{Models and numerical procedure} \label{models}

\subsection{The models}

We restrict ourselves to axisymmetric models. Mean fields may well be defined by azimuthal averages. Also, the large-scale solar magnetic field is largely axisymmetric; the  deviations from axisymmetry, as they appear, for instance, in the form of active longitudes, are in general small. Nevertheless, a more comprehensive study would have to check whether our {\em model} is stable against non-axisymmetric
perturbations.

The axisymmetric LSMF is represented  in the form
\begin{equation}
\left\langle \vec{B}\right\rangle =\mathrm{curl}\left( \frac{A\vec{e}_{\phi}}{r\sin\theta}\right) +B\,\vec{e}_{\phi}
\end{equation}
as the sum of a poloidal and a toroidal component;
$A(r,\theta,t)$ and $B(r,\theta,t)$ are scalar functions of 
radius $r$, colatitude $\theta$ and time $t$, and  $\vec{e}_{\phi}$ is
the unit vector in the direction of the azimuthal coordinate $\phi$. The mean-field induction equation,
containing the effects of differential rotation, expressed by the rotation
rate $\Omega(r,\theta)=|\vec{\Omega}(r,\theta)|$, and of the MEMF, $\vec{\mathcal{E}}$, then takes the form
\begin{align}
\frac{\partial A}{\partial t} & = r\sin\theta\,\mathcal{E}_{\phi} \,,\label{eq:2} \\
\frac{\partial B}{\partial t} & = \frac{1}{r}\frac{\partial\left(\Omega,A\right)}{\partial\left(r,\theta\right)}+\frac{1}{r}\left(\frac{\partial r\mathcal{E}_{\theta}}{\partial r}-\frac{\partial\mathcal{E}_{r}}{\partial\theta}\right) \,.\label{eq:1}
\end{align}

\subsubsection{The solar $\alpha^2\delta\Omega$ dynamo}
\label{alpha_delta_Omega}

In the solar $\alpha^2\delta\Omega$ dynamo model, to keep the numerical effort
(i.e., the spectral resolution, see Sect.~\ref{sec_numerics} below) at a minimum,
only the azimuthal
$\vec{\Omega}\times\vec{J}$ effect is taken into account. This may be justified by the fact that
the toroidal part of the solar LSMF is much stronger than the poloidal one. However,
the other parts of the MEMF ($\alpha$ effect, isotropic and anisotropic turbulent diffusion, turbulent pumping)
are included in all components.
Using the results of \citet{pip08},
the components of the
MEMF in spherical coordinates become
\begin{equation}
\begin{split}
\mathcal{E}_{r} ={} &
\tilde{\eta}_{T}\left\{
-\frac{f_{2}^{(d)}+\left(1+\varepsilon\right)f_{1}^{(a)}\sin^{2}\theta}{r\sin\theta}\frac{\partial\sin\theta
    \,B}{\partial\theta}\right.\\
&{}-\frac{\left(1+\varepsilon\right)f_{1}^{(a)}\sin2\theta}{2r}\frac{\partial
    rB}{\partial r}-G\sin2\theta\,
f_{1}^{(a)}\, B\\
&{}+ C_{\alpha}\left[ G\left(f_{5}^{(a)}\cos^{2}\theta+f_{10}^{(a)}+2f_{6}^{(a)}\right)\frac{\cos\theta}{r^{2}\sin\theta}\frac{\partial A}{\partial\theta}\right. \\
&{}+ U\left(f_{4}^{(a)}\cos^{2}\theta+f_{11}^{(a)}+2f_{8}^{(a)}\right)\frac{\cos\theta}{r^{2}\sin\theta}\frac{\partial A}{\partial\theta} \\
&{}+ 
 \left(f_{5}^{(a)}\cos^{2}\theta+f_{6}^{(a)}-f_{7}^{(a)}\right)\frac{G}{r}\frac{\partial
   A}{\partial r} \\
&{}+\left.\left.\left(f_{4}^{(a)}\cos^{2}\theta+f_{8}^{(a)}-f_{9}^{(a)}\right)\frac{U}{r}\frac{\partial
  A}{\partial r}\right]\right\} \,,\label{eq:er}
\end{split}
\end{equation}
\begin{equation}\begin{split}
\mathcal{E}_{\theta} ={} &
\tilde{\eta}_{T}\left\{
\frac{f_{2}^{(d)}+\left(1+\varepsilon\right)
f_{1}^{(a)}\cos^{2}\theta}{r}\frac{\partial
    rB}{\partial r} \right.  \\
&{}-\frac{\left(1+\varepsilon\right)f_{1}^{(a)}\cos\theta}{r}\frac{\partial\sin\theta
    \,B}
{\partial\theta}\\
&{}-
 \left[Gf_{3}^{(a)}+\left(\varepsilon-1\right)Uf_{2}^{(a)}\right. \\
&{}+\left. G(\cos^{2}\theta-\sin^{2}\theta)f_{1}^{(a)}\right]B \\
&{}-C_{\alpha}\left[
\left(f_{5}^{(a)}\sin^{2}\theta+f_{10}^{(a)}\right)\frac{G\cos\theta}{\sin\theta}\frac{\partial
   A}{\partial r} \right.\\
&{}+
\left(f_{5}^{(a)}\cos^{2}\theta+f_{6}^{(a)}+f_{7}^{(a)}\right)\frac{G}{r}
\frac{\partial A}{\partial\theta} \\
 &{} +\left(f_{4}^{(a)}\sin^{2}\theta+f_{11}^{(a)}\right)\frac{U\cos\theta}{\sin\theta}\frac{\partial
   A}{\partial r} \\
&{}+\left.\left.\left(f_{4}^{(a)}\cos^{2}\theta+f_{8}^{(a)}+f_{9}^{(a)}\right)\frac{U}{r}\frac{\partial A}{\partial\theta}\right]\right\} \,,\label{eq:et}
\end{split}\end{equation}
\begin{equation}\begin{split}
\mathcal{E}_{\phi} = {}&
\frac{\tilde{\eta}_{T}}{r\sin\theta}\Bigg\{
\left[f_{2}^{(d)}+f_{1}^{(a)}\Big((1+\varepsilon)\right.\\
&{}+\left. \left.\left.(\varepsilon-1)\sin^{2}\theta\right)\right]\frac{\partial^{2}A}
{\partial r^{2}} +  \left[f_{2}^{(d)}+ f_{1}^{(a)}\Big(2\varepsilon \right. \right. \\
&{}+\left. \left. (1-\varepsilon)\sin^{2}\theta\right)\right]\frac{\sin\theta}{r^{2}}\frac{\partial}{\partial\theta}\frac{1}{\sin\theta}\frac{\partial A}{\partial\theta} \\
 &{} +
 \frac{\left(1-\varepsilon\right)f_{1}^{(a)}}{r}\left(\frac{3\sin2\theta}{2r}\frac{\partial
     A}{\partial\theta}-\sin2\theta\frac{\partial^{2}A}{\partial
     r\partial\theta}\right. \\
&{}+\left.\sin^{2}\theta\frac{\partial A}{\partial r}\right)
 +\sin2\theta f_{1}^{(a)}\left[\left(\varepsilon-1\right)U+\varepsilon G\right]\frac{\partial A}{\partial\theta} \\
 &{} -
 \left[f_{1}^{(a)}\left(G-\left(G\varepsilon+\left(\varepsilon-1\right)U\right)\sin^{2}\theta\right)\right.
 \\
&{}+\left. Gf_{3}^{(a)}+\left(\varepsilon-1\right)Uf_{2}^{(a)}\right]\frac{\partial
  A}{\partial r} \\
&{}+ C_{\alpha}Br\sin2\theta Gf_{12}^{(a)}
 \\
&{} + \left.
C_{\delta}\varepsilon f_{4}^{(d)}\left(\frac{r\sin2\theta}{2}\frac{\partial B}{\partial r}-\sin^{2}\theta\frac{\partial B}{\partial\theta}\right)\right\} \,.\label{eq:ef}
\end{split}\end{equation}
Here $G=(\partial/\partial r)\log\rho$ and $U=(\partial/\partial r)\log\left(u_{c}^{2}\right)$ are the scale factors of density ($\rho$) and turbulence intensity, respectively, and
 $\tilde{\eta}_{T}=C_{\eta}\,\eta_T$, with
$\eta_T=u_{c}\ell_{c}/3$, where $\ell_{c}$ denotes the correlation length of the background
turbulence.
$C_{\eta}\leq1$,
 $C_{\alpha}\leq1$, $C_{\delta}\leq1$
are parameters to control the relative strengths of different turbulence effects.
$C_{\alpha}$ and $C_{\delta}$ weight the $\alpha$ effect and the
 $\vec{\Omega}\times\vec{J}$ effect, respectively,
and the Prandtl-like number $C_{\eta}$ regulates the turbulence level, with which all  contributions are equally scaled.
Below, we consider the case $C_{\eta}=1/5$.

A remark concerning the introduction of weighting factors for the $\alpha$ and $\vec{\Omega}\times\vec{J}$ effects seems to be in order. First, we wish to study the dynamo
onset. For the conditions of the Sun, the two effects have therefore to be reduced in their strength.
Second, the $\tau$ approximation, which is used to calculate $\vec{\mathcal{E}}$, is based
on heuristic closure assumptions for the turbulence; for a critical analysis of the approximation
we refer to \citet{radrhe07}. Therefore, we have left a freedom to adjust
at least the relative strengths of different turbulence effects.

The integration
domain is radially bounded by $r=0.72R_{\odot}$ and $r=0.96R_{\odot}$,
where the boundary conditions are
$A=0$, ${\displaystyle\frac{\partial rB}{\partial r}=0}$ at the bottom boundary
\citep[a usual approximation to perfect-conductor conditions, see, e.g.,][]{koh73,jouetal08},
and vacuum conditions (that is, $B=0$ and continuous match
of the poloidal field component to an exterior potential field) at the top boundary.

In our numerical calculations we have used a dimensionless
form of 
the equations,
 substituting $r=xR_{\odot}$
and $t\rightarrow\eta_{0}t/R_{\odot}^{2}$,
where $\eta_0$  is the maximum value of $\eta_T$ in the convection zone. The estimated value for
$\eta_0$ is $\sim10^{9} \mathrm{m}^2/\mathrm{s}$. The turbulent magnetic diffusion time on the basis
of this value and the solar radius, our time unit, is about $15\,\mathrm{yr}$. Thus, in order to match the
solar conditions, the models should give cycle periods of the order of the turbulent diffusion time.

For the construction of the model the current knowledge of the rotation rate in the convection zone \citep{schouetal98} was taken into account. In the numerical calculations,
the rotation profile  was
approximated by \citep[cf., e.g.,][]{godroz00}
\begin{equation}
 \Omega(x,\theta)=\Omega_{0}\,f(x,\theta)
\end{equation}
with 
\begin{equation}\begin{split}
f\left(x,\theta\right)={}&\frac{1}{435}\left[435+50\left(x-x_{0}\right)
+22\phi\left(x\right)\left(1-5\cos^{2}\theta\right)\right.\\
&{}-\left.3.5\left(1-14\cos^{2}\theta+21\cos^{4}\theta\right)\right],
\end{split}\end{equation}
where
\begin{equation}
\phi(x)=0.5[1+\mathrm{erf}(50(x-x_{0}))]
\end{equation}
and $x_{0}=0.71$ is the position of tachocline, situated below the bottom boundary
of the integration
domain.

The radial profiles of characteristic
quantities of the turbulence, such as the rms value $u_{c}$
and the
correlation length and time $\ell_{c}$ and $\tau_{c}$ of the
convective background velocity field, as well as the density stratification
were calculated
on the basis of a standard model of the solar interior \citep{sti02}.

\begin{figure*}
\centering
\includegraphics[width=0.5\linewidth]{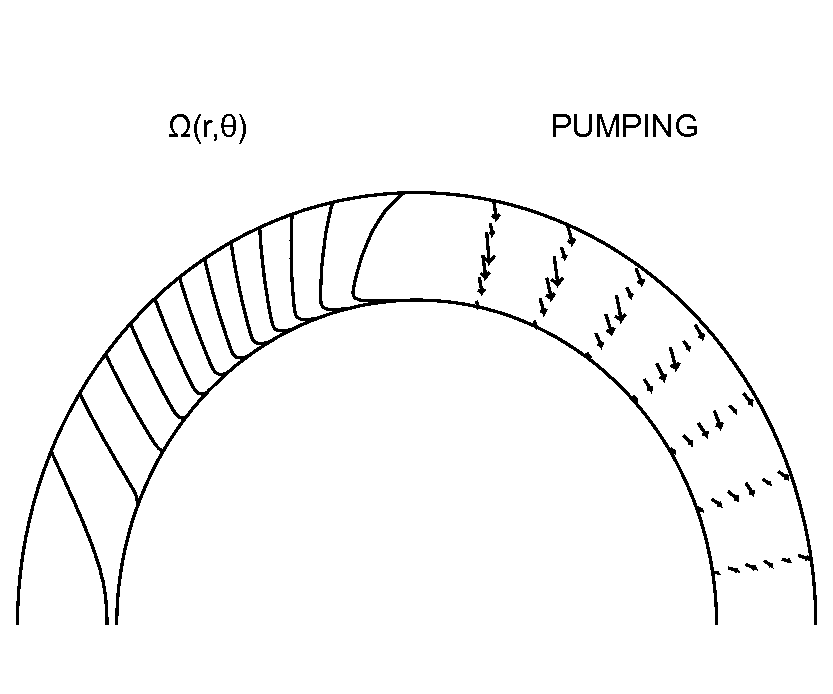}\includegraphics[width=0.5\linewidth]{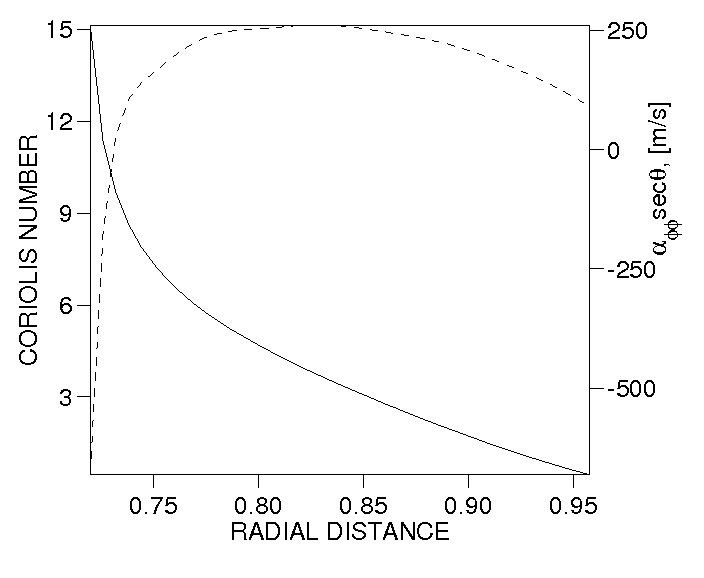}
\caption{\label{fig:fig0}
Basic model quantities. Left panel: Contours of the rotation rate
in the solar convection zone (left) and
geometry of the pumping velocity for the toroidal part of the LSMF (right). Right panel: Radial profiles of the Coriolis
number $\Omega^{*}$ (solid line)
and  $\alpha_{\phi\phi}\sec\theta$ (dashed line).
$\alpha_{\phi\phi}$ changes sign near the bottom
of the convection zone.
}
\end{figure*}

Some basic
model quantities, namely, the differential rotation,
the turbulent pumping velocity for the toroidal component of the LSMF and
the radial profiles of the Coriolis number $\Omega^{*}$
and  of $\alpha_{\phi\phi}\sec\theta$ ($\alpha_{\phi\phi}$ measures the strength of the azimuthal $\alpha$ effect) are shown in Fig.~\ref{fig:fig0} \citep[see also][]{seeetal03,kuzpipsee06}. 

\subsubsection{The $\delta^{2}$ dynamo}

In our $\delta^{2}$ dynamo model for a full sphere, representing a star, differential rotation and the $\alpha$ effect are neglected.
For simplicity, we assume a constant density
stratification $R_{\star}G=-20$  throughout the star ($R_{\star}$
is the radius of the star),
corresponding to the value of $G$ at a distance of $R_{\star}/2$ from the center of an M5 dwarf
with mass $M=0.2M_{\odot}$, luminosity $L=0.0058L_{\odot}$, radius $R_{\star}=0.23R_{\odot}$ and surface temperature $T_{\mathrm{surf}}=3458\,\mathrm{K}$
(we used the stellar evolution code TWIN \citep{egg71,eggkis02} in the frontend-version WTTS \citep{izzgle06}).
We also neglect the effects of the spatial inhomogeneity of the turbulence
and set $\tilde{\eta}_{T}=\eta_0$, with $\eta_0$ denoting the value of the homogeneous turbulent magnetic
diffusivity.
Furthermore, we set $C_{\delta}=1$. At the surface vacuum boundary conditions are imposed.

Different from the treatment of the solar $\alpha^2\delta\Omega$ dynamo model in Sect.~\ref{alpha_delta_Omega}, now also the contributions of the $\vec{\Omega}\times\vec{J}$ effect
to the $r$ and $\theta$ components of $\vec{\mathcal{E}}$ are taken into account. The equations for
the poloidal potential $A$ and for $\mathcal{E}_{\phi}$, Eqs. (\ref{eq:2}) and (\ref{eq:ef}), remain unchanged (except for the omission of the term proportional to $C_{\alpha}$ and $U=0$
in Eq.~(\ref{eq:ef})),
while in the equation for the toroidal potential $B$, Eq.~(\ref{eq:1}), the first term on the right-hand side, describing
the effect of the differential rotation, vanishes. 
We write the equation for $B$ in the form
\begin{equation}
\frac{\partial B}{\partial t}=\frac{1}{r}\left(\frac{\partial r\left(\mathcal{E}_{\theta}+\mathcal{E}^{\left(\delta\right)}_{\theta}\right)}{\partial r}-\frac{\partial\left(\mathcal{E}_{r}+\mathcal{E}^{\left(\delta\right)}_r\right)}{\partial\theta}\right),\label{eq:dlt1}\end{equation}
where $\mathcal{E}_r$ and $\mathcal{E}_{\theta}$ are given by
Eqs. (\ref{eq:er}) and (\ref{eq:et}) in Sect.~\ref{alpha_delta_Omega}
(with $\tilde{\eta}_{T}=\eta_0$, $U=0$, $C_{\alpha}=0$)
and  $\mathcal{E}^{(\delta)}_r$ and $\mathcal{E}^{(\delta)}_{\theta}$ denote the additional contributions due to the
$\vec{\Omega}\times\vec{J}$ effect. For these we have
\begin{equation}\begin{split}
\mathcal{E}^{\left(\delta\right)}_r  ={} &
\eta_0\varepsilon\left\{\frac{\cos\theta\left(c_{3}f_{3}^{(d)}\cos2\theta+f_{4}^{(d)}\right)}{r^{2}\sin\theta}\left(\frac{\partial^{2}A}{\partial
    r\partial\theta}-\frac{1}{r}\frac{\partial
    A}{\partial\theta}\right) \right. \\
&{}+\frac{c_{3}f_{3}^{(d)}\cos^{2}\theta}{r}\frac{\partial^{2}A}{\partial r{}^{2}}\\
 &{} -
\left. \frac{\left(c_{3}f_{3}^{(d)}\cos^{2}\theta+f_{4}^{(d)}\right)}{r^{2}}\left(\frac{1}{r}\frac{\partial^{2}A}{\partial\theta^{2}}+\frac{\partial
     A}{\partial r}\right)\right\}, \label{eq:er-dlt2}
\end{split}\end{equation}
\begin{equation}\begin{split}
\mathcal{E}^{\left(\delta\right)}_{\theta} ={} &
\eta_0\varepsilon\left\{\frac{\left(f_{4}^{(d)}-c_{3}f_{3}^{(d)}\cos2\theta\right)}{r^2}\left(\frac{\partial^{2}A}{\partial
    r\partial\theta}-\frac{1}{r}\frac{\partial
    A}{\partial\theta}\right)\right\} \\
&{}+\frac{\sin2\theta}{2r^2}c_{3}f_{3}^{(d)}\left(\frac{1}{r}\frac{\partial^{2}A}{\partial\theta^{2}}+\frac{\partial A}{\partial r}\right)\\
 &{} - \left. \frac{\cos\theta\left(f_{4}^{(d)}+c_{3}f_{3}^{(d)}\sin^{2}\theta\right)}{r\sin\theta}\frac{\partial^{2}A}{\partial r^{2}}\right\} \,.\label{eq:et-dlt2}
\end{split}\end{equation}
$c_{3}$ is a weighting factor for the  component of the $\vec{\Omega}\times\vec{J}$
effect with coefficient $f_3^{(d)}$ (i.e., the component of third order in $\vec{e}$). There is no dynamo with $c_{3}=1$. The meaning of this coefficient
for the model will be explained further in Sect.~\ref{sub:d2} below. Dynamo action sets in
when both $c_{3}$ and the Coriolis number exceed  threshold values.
In our numerical calculations for this model, length and time are normalized to $R_{\star}$ and
$R_{\star}^2/\eta_0$, respectively.

\subsection{The numerical procedure} \label{sec_numerics}

The eigenvalue problem is treated by means of a Galerkin method.
We seek the solutions to Eqs.~(\ref{eq:2}--\ref{eq:1}) for the $\alpha^{2}\delta\Omega$
dynamo and to Eqs.~(\ref{eq:2},\ref{eq:dlt1})  for the $\delta^{2}$ dynamo
in the form
 \begin{align}
A\left(x,\theta,t\right) & =  \mathrm{e}^{{\displaystyle
      \lambda t}}\sum_{n}\sum_{m}a_{nm}
\sin\theta \, S_{nm}^{(A)}\left(x\right)P_{m}^{1}\left(\cos\theta\right),\label{eq:A-dec} \\
B\left(x,\theta,t\right) & =  \mathrm{e}^{\displaystyle{\lambda t}}\sum_{n}\sum_{m}b_{nm}S_{n}^{(B)}\left(x\right)P_{m}^{1}\left(\cos\theta\right),\label{eq:B-dec}
\end{align}
where $S_{nm}^{(A)}$ and $S_{n}^{(B)}$ are linear combinations of Legendre polynomials and
 $P_{m}^{1}$ is the associated Legendre function of degree $m$ and order $1$. By these expansions
the regularity of the solutions at the poles
$\theta=0$ and $\theta=\pi$, where both $A$ and $B$ are set to zero, is ensured.
To satisfy the conditions at the radial boundaries,
we use the {}``basis recombination'' of the Legendre polynomials
\citep[see][]{boy01}.
In the case of the solar $\alpha^{2}\delta\Omega$ dynamo
this reads
\begin{align}
S_{nm}^{(A)}\left(x\right)& =  P_{n}\left(x\right)+a_{1}P_{n+1}\left(x\right)+a_{2}P_{n+2}\left(x\right),\label{eq:A-dec1}\\
S_{n}^{(B)}\left(x\right) & =  P_{n}\left(x\right)+b_{1}P_{n+1}\left(x\right)+b_{2}P_{n+2}\left(x\right),\label{eq:B-dec1}
\end{align}
where $P_{n}(x)$ is the Legendre polynomial of degree $n$ and
\begin{align}
a_{1}  =  \frac{2n+3}{(n+2)^{2}+2m}\,, &\quad a_{2}=-\frac{(n+1)^{2}+2m}{(n+2)^{2}+2m} \,, \\
b_{1}  =  -\frac{4n+4}{2n^2+6n+3}\,, &\quad b_{2}=-\frac{2n^2+2n-1}{2n^2+6n+3} \,.
\end{align}
For the $\delta^{2}$ dynamo we exploit the symmetry of the problem,
as $(x,\theta)$ and $(-x,\theta+\pi)$ represent the same point.
The boundary conditions on $x$ are then satisfied with
\begin{align}
S_{nm}^{(A)}\left(x\right)& = 
x\left(P_{n}\left(x\right)
-\frac{2+2m+n(n+1)}{2+2m+(n-2)(n-1)}P_{n-2}\left(x\right)\right),\label{eq:A-dec2} \\
S_{n}^{(B)}\left(x\right)& =  x\left(P_{n}\left(x\right)-P_{n-2}\left(x\right)\right),\label{eq:B-dec2}
\end{align}
and the summations in Eqs.~(\ref{eq:A-dec}--\ref{eq:B-dec}) run over even $m+n$ only
and start from $n=2$.

Integrations over radius and latitude, necessary to calculate the expansion coefficients
$a_{nm}$ and $b_{nm}$, were done by means of the Gauss-Legendre
procedure. The used computer code was developed employing the free computer algebra system
Maxima, and the eigenvalue problem was
solved with the help of Lapack routines, which are accessible within Maxima.
For the solar dynamo problem we used the first 7 modes in the 
radial expansion and the first 20 modes in the latitudinal expansion, while in the case of the $\delta^{2}$
dynamo a decomposition with $16\times16$ modes was applied. The results
were confirmed by a number of runs with higher resolutions.

\section{Results} \label{results}

\subsection{The solar $\alpha^{2}\delta\Omega$ dynamo}

\subsubsection{Effects of anisotropic diffusion and the small-scale dynamo in the $\alpha^{2}\Omega$
dynamo}

From Eq.~(\ref{dif1}) or Eqs.~(\ref{eq:er}--\ref{eq:ef}) it is seen that for the case of energy
equipartition, $\varepsilon=1$, the small-scale dynamo makes significant
contributions to different parts of the MEMF.
\citet{kit02,kit04} has found that anisotropic diffusion may be in a large part responsible
for the equatorial drift of the toroidal LSMF.
This drift is actually an extra diffusion (of both the toroidal and poloidal parts of the
mean field)
along the rotation axis $\vec{e}$.
In Eq.~(\ref{dif1}) it is described by the
second part of the anisotropic-diffusion term, which is proportional
to $f_1^{(a)}(\varepsilon+1)$. Thus, for $\varepsilon=1$ the drift is enhanced by a factor of two compared to the case of $\varepsilon=0$.
The first part of the anisotropic-diffusion term in Eq.~(\ref{dif1}) can in vector form be written as
 $2f_1^{(a)}\varepsilon\,\vec{e}\times\nabla(\vec{e}\cdot\langle\vec{B}\rangle)$. Due to the assumed
axisymmetry, it is purely toroidal (azimuthal). Therefore, it does not influence the toroidal part of the
mean field. It actually describes a diffusion of the poloidal part of the mean field in the direction
perpendicular to the rotation vector $\vec{e}$.

\begin{figure*}
\centering
\includegraphics[width=0.9\linewidth]{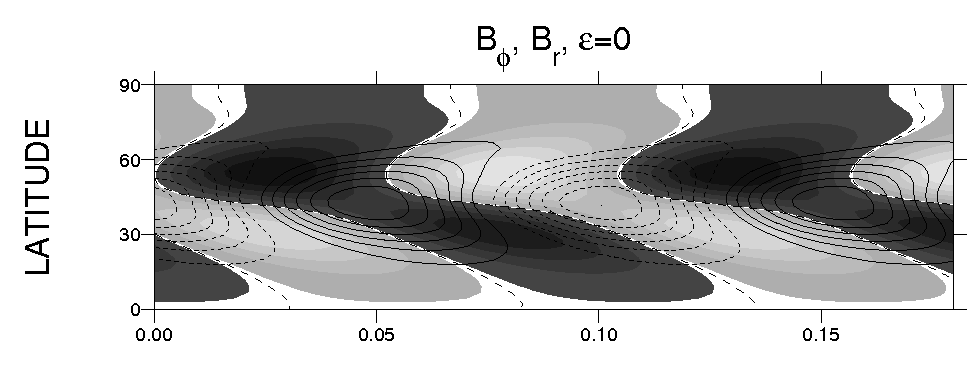}
\includegraphics[width=0.9\linewidth]{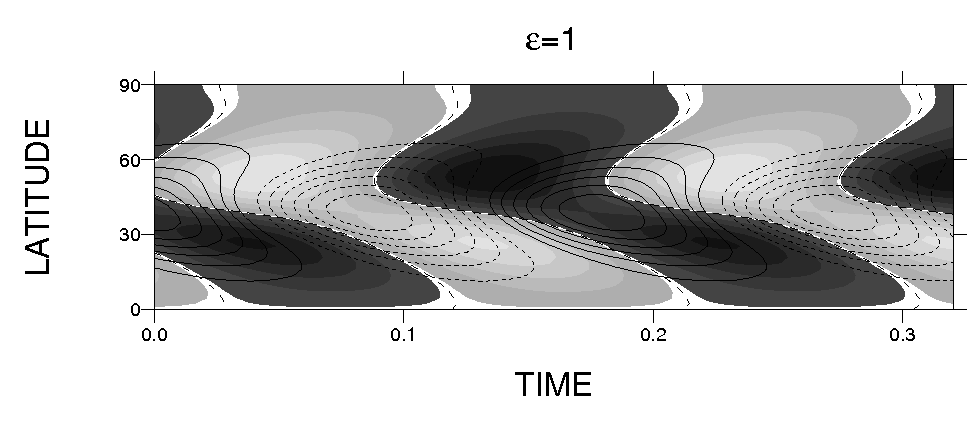}
\includegraphics[width=0.9\linewidth]{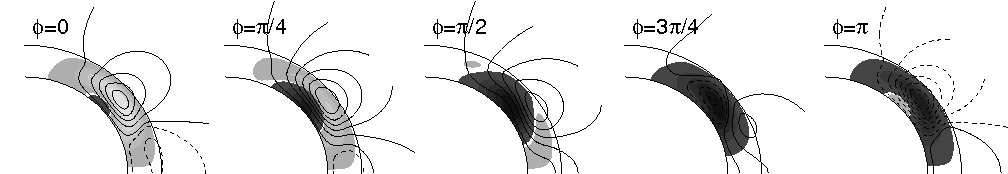}
\caption{\label{fig:Effects-of-E}Effects of  anisotropic
diffusion and the small-scale dynamo for an
 $\alpha^{2}\Omega$ dynamo model on the basis of the first unstable dipolar eigenmode
($C_{\alpha}=0.087$). Top and middle: Butterfly diagrams in the form of 
isocontours of the toroidal LSMF (integrated over depth
in the convection zone) and overlaid greyscale plots of the radial
LSMF at the top boundary
 without (top) and with (middle) the effects of anisotropic diffusion
and the small-scale dynamo.
 Time is measured in units of $R_{\odot}^{2}/\eta_0$.
Solid/dashed lines and bright/dark areas indicate positive/negative field values.
 Bottom: Snapshots of the strength of the
toroidal LSMF (greyscale plot) and field lines of the poloidal LSMF
over half a cycle for the model containing the effects of anisotropic diffusion
and the small-scale dynamo. Solid/dashed lines indicate
clockwise/counter-clockwise field direction along the poloidal field lines.
}
\end{figure*}

In Fig.~\ref{fig:Effects-of-E} (top and middle),
showing butterfly diagrams in the form of 
isocontours of the toroidal LSMF (integrated over depth
in the convection zone) and overlaid greyscale plots of the radial
LSMF at the top boundary,
the effects of anisotropic diffusion and the small-scale dynamo are demonstrated for a pure  $\alpha^{2}\Omega$ dynamo model on the basis of the first unstable dipolar eigenmode.
By dipolar/quadrupolar modes we mean modes that are antisymmetric/symmetric with respect
to the equatorial plane.
The used dipolar mode  is not the primary dynamo eigenmode, cf. Fig.~\ref{fig:fig1} (left) in Sect.~\ref{sec_alpha_delta_Omega} below.
Qualitatively, the results presented here resemble those of
\citet{kit02}, who did not include the effect of  the small-scale
dynamo and used a slightly different rotation law.

As is seen in Fig.~\ref{fig:fig0} (left), the direction of the turbulent pumping velocity is such that
it transports magnetic flux toward the equator. So also the pumping effect offers a possibility
to explain the observed latitudinal distribution and drift of the solar activity phenomena
\citep{guegou08}. In our model the pumping effect is obviously to weak to produce a noticeable
equatorward drift of the toroidal LSMF.

Finally, as is seen in Fig.~\ref{fig:Effects-of-E} (bottom), the dynamo wave moves
mainly radially outward.
This is in accordance with the tendency of the $\alpha\Omega$ dynamo waves to move along  isorotation surfaces \citep{yos75}. In the bulk of the convection zone these surfaces are largely parallel to  radius (cf. Fig. \ref{fig:fig0}, left panel). A better agreement with the observations could thus
be expected if the toroidal LSMF were confined to the tachocline, where the rotation rate varies mainly with radius. This would then amount to a boundary-layer dynamo, which is
basically different from the
distributed convection-zone dynamo considered here. A more complete model of the solar dynamo would include the dynamo effects of both the tachocline and the convection zone proper.

\subsubsection{The $\alpha^{2}\delta\Omega$ dynamo} \label{sec_alpha_delta_Omega}

\begin{figure*}
\includegraphics[width=0.45\linewidth]{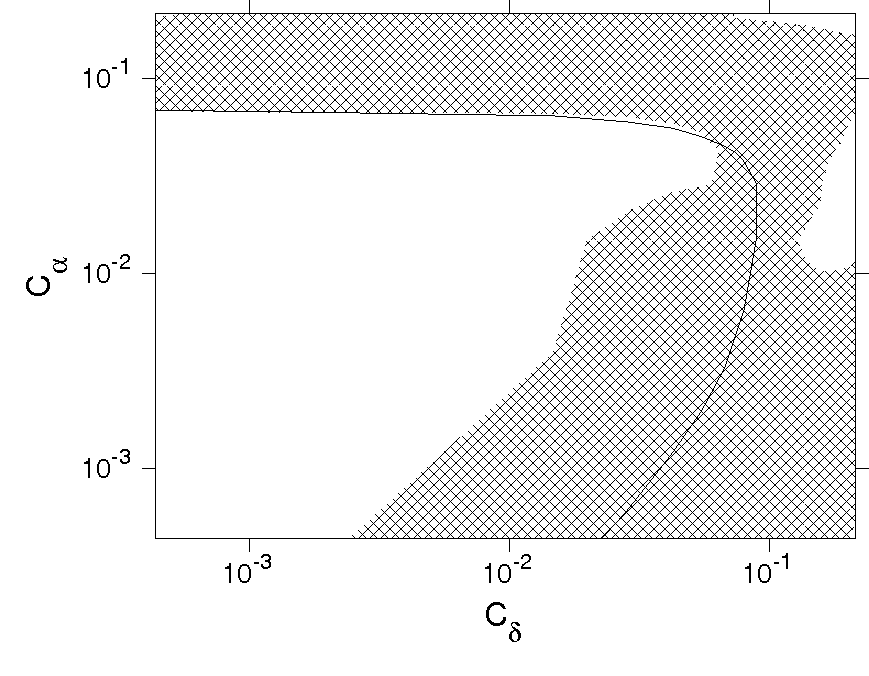} ~~\includegraphics[width=0.5\linewidth]{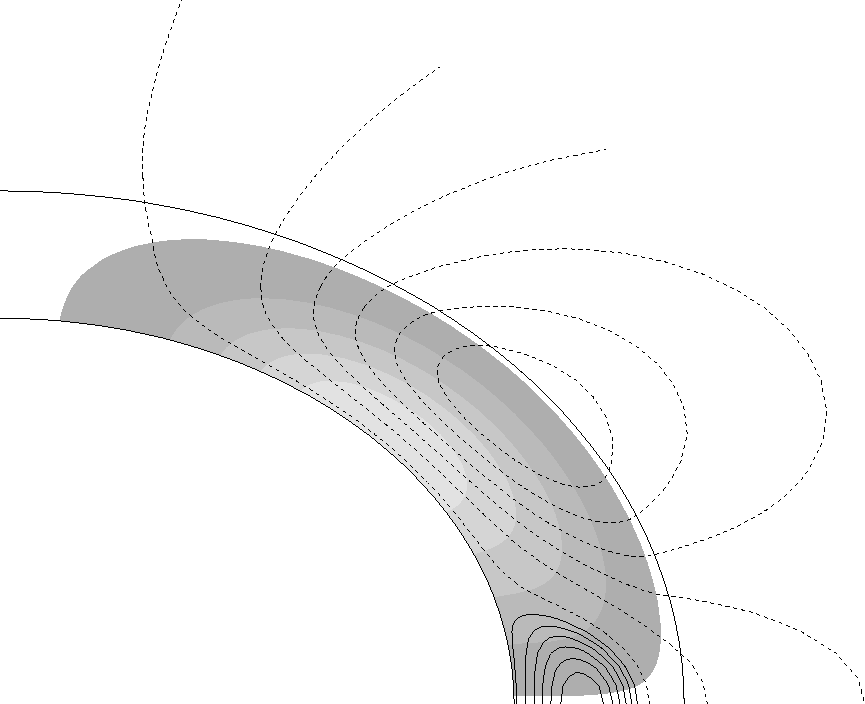}
\caption{\label{fig:fig1} Left: Stability boundary for the $\alpha^{2}\delta\Omega$ dynamo
in the $C_{\delta}$-$C_{\alpha}$ plane.
The stable region (no dynamo) lies below and to the left of the solid line. Shading indicates dominance
of dipolar modes. Right: First unstable
eigenmode for the steady $\delta\Omega$ dynamo. A greyscale plot of the toroidal
LSMF and field lines of the poloidal LSMF are shown. Solid/dashed lines indicate
clockwise/counter-clockwise field direction along the poloidal field lines.}
\end{figure*}

Figure \ref{fig:fig1} (left)
shows a linear-stability diagram for the $\alpha^{2}\delta\Omega$ dynamo, i.e., the stability boundary for the most unstable eigenmode
in the $C_{\delta}$-$C_{\alpha}$ plane.
The $\alpha^{2}\Omega$ dynamo is included as a limiting case ($C_{\delta}=0$). In the figure, regions where the
eigenmode with the largest growth rate is a dipolar mode are indicated by shading.
In the limiting case of $C_{\delta}=0$ the dipolar modes  are not the most unstable modes. Though
a dipolar mode becomes unstable only slightly above the total-stability boundary,
there is a preference for
quadrupolar modes in the pure $\alpha^{2}\Omega$ dynamo, contrary to the antisymmetric parity characterizing the large-scale solar magnetic field.
The parity issue
is met in other dynamo models, like, e.g., the
flux-transport models, as well \citep[see, e.g.,][]{diketal05}. For
the pure $\delta\Omega$ dynamo ($\vec{\Omega}\times\vec{J}$ effect plus differential rotation),
on the other hand,
there is a preference of dipolar modes ($C_{\alpha}=0$ in Fig.~\ref{fig:fig1}). These are, however, non-oscillatory.
An example of such a mode is shown in Fig.~\ref{fig:fig1} (right). 
The fact that in the pure $\delta\Omega$ model only non-oscillatory modes are found seems to indicate that for models of the solar dynamo the
$\alpha$ effect is also needed, at least for models without
meridional circulation.

\begin{figure}
\centering
\includegraphics[angle=0,width=1\columnwidth]{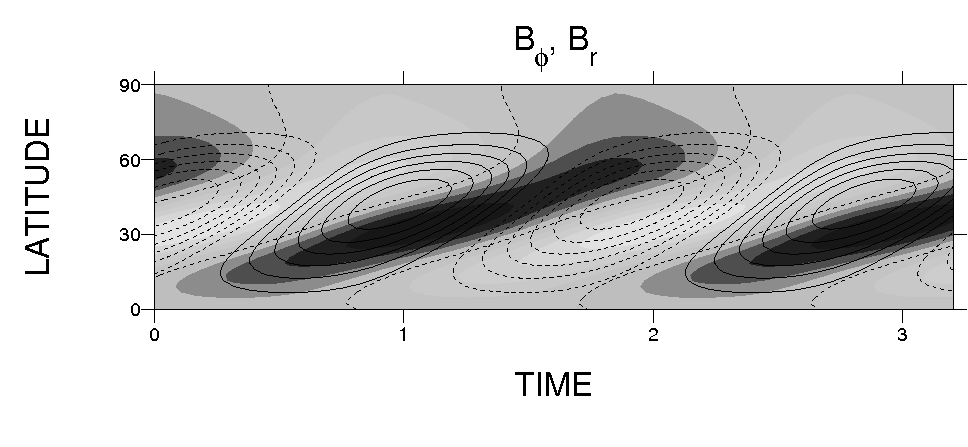}
\caption{\label{fig:fig2}Butterfly diagram in the form of 
isocontours of the toroidal LSMF (integrated over depth
in the convection zone) and overlaid greyscale plot of the radial
LSMF at the top boundary for an $\alpha^2\delta\Omega$ dynamo with
 $C_{\delta}=0.07$, $C_{\alpha}=0.01$.
Time is measured in units of $R_{\odot}^{2}/\eta_0$.
Solid/dashed lines and bright/dark areas indicate positive/negative field values.}
\end{figure}

For increasing $C_{\alpha}$, oscillatory dipolar modes are excited if $C_{\alpha}$ is no longer very small
compared to $C_{\delta}$. The properties of these modes are determined by the ratio
$C_{\alpha}/C_{\delta}$.
Fig.~\ref{fig:fig2}
 shows a
simulated butterfly diagram (isocontours of the toroidal LSMF with an overlaid greyscale
plot of the radial LSMF at the top boundary) for $C_{\delta}=0.07$, $C_{\alpha}=0.01$. 
For these parameter values we obtain
a slow dynamo wave propagating from the equator to the pole for both the
toroidal and poloidal fields, in contrast to the solar observations. The period of the dynamo is considerably
increased compared to the pure $\alpha^{2}\Omega$ dynamo shown in
Fig.~\ref{fig:Effects-of-E}. 

\begin{figure*}
\centering
\includegraphics[angle=0,width=12cm]{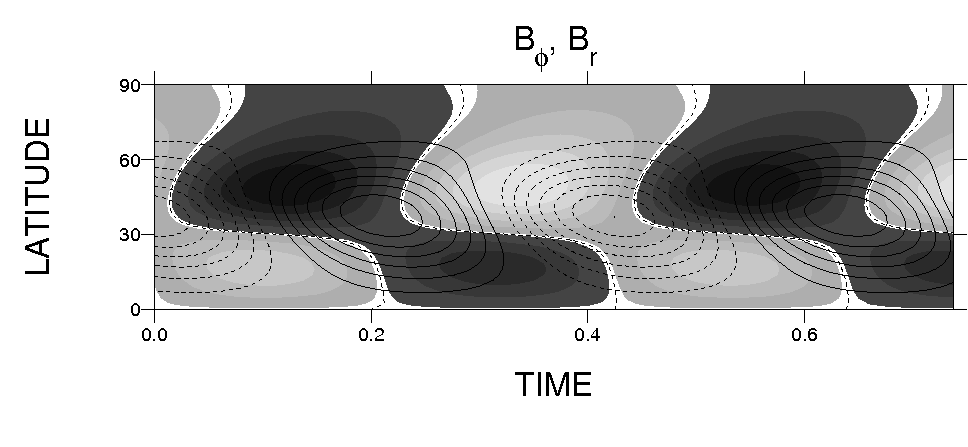}
\includegraphics[angle=0,width=12cm]{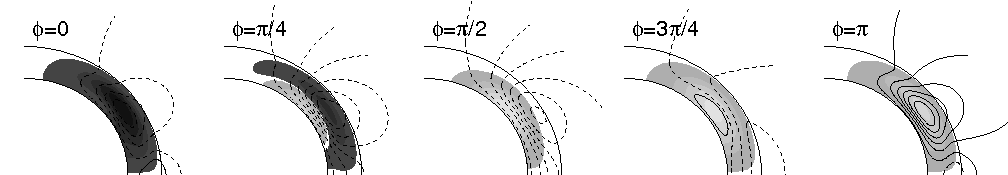}
\caption{\label{fig:fig3}Top: Butterfly diagram in the form of 
isocontours of the toroidal LSMF (integrated over depth
in the convection zone) and overlaid greyscale plot of the radial
LSMF at the top boundary for an $\alpha^2\delta\Omega$ dynamo with
 $C_{\delta}=0.1$, $C_{\alpha}=0.05$.
Time is measured in units of $R_{\odot}^{2}/\eta_0$.
Solid/dashed lines and bright/dark areas indicate positive/negative field values.
Bottom:
Snapshots of the strength of the
toroidal LSMF (greyscale plot) and field lines of the poloidal LSMF
over half a cycle. Solid/dashed lines indicate
clockwise/counter-clockwise field direction along the poloidal field lines.}
\end{figure*}

In the intermediate case, when the $\vec{\Omega}\times \vec{J}$
effect exceeds the $\alpha$ effect while the latter one is not very small compared to the first one,
we find a good match with the solar case. A corresponding butterfly diagam
and snapshots of the toroidal and poloidal fields over half a cycle
for $C_{\delta}=0.1$, $C_{\alpha}=0.05$ are shown in
Fig.~\ref{fig:fig3}.
While the wings of the obtained dynamo waves are too wide, other qualitative
properties, such as the directions of the drifts of the toroidal and poloidal parts
of the magnetic field and their phase relation
(e.g., polar reversal of $B_{r}$ shortly after maximum
of $B_{\phi}$ at low latitudes) are in good agreement with the observations.
The period of the dynamo is shorter than but comparable with the turbulent magnetic diffusion
time. In the range $1<C_{\delta}/C_{\alpha}<3$ the match with the solar observations is best (it can be
seen in Fig.~\ref{fig:fig1} that in this range the dipolar modes are dominating at the stability
boundary).

The obtained ratio $B_\mathrm{T}/B_\mathrm{P}$ of the toroidal ($B_\mathrm{T}$) and poloidal ($B_\mathrm{P}$) field strengths is about 50, the ratio of the toroidal and poloidal field energies thus of the order of $10^3$. This comes close to the
ratio $B_\mathrm{T}/B_\mathrm{P}\sim 100$ estimated from measurements
of the fields in active regions [yielding the estimate $B_\mathrm{T}\sim 200\,\mathrm{G}$ for a distributed toroidal field in the convection zone, cf. \citet[][Sect. 8.4.1]{sti02}] and the field strength at the solar poles (yielding the estimate $B_\mathrm{P}\sim 1\,\mathrm{G}$ for the poloidal field).

\subsection{\label{sub:d2} The $\delta^{2}$ dynamo}

The usual $\vec{\Omega}\times\vec{J}$ effect, given by the term
$-\varepsilon f_4^{(d)}\vec{e}\times\mu_0\vec{J}$ in the expression for the turbulent emf (cf. Eq.~(\ref{first-order-terms})), does not contribute to $\vec{\mathcal{E}}\cdot\vec{J}$.
 Therefore,
it cannot bring energy into the mean magnetic field and is not capable of dynamo action
when working alone. It can yield a dynamo, however, when acting together with differential rotation. The role
of the effect is then to transfer energy from the toroidal to the poloidal field. 
In our model of the $\delta^{2}$ dynamo differential rotation is neglected
(as well as the $\alpha$ effect). Energy has thus to be provided by the additional source term
in our model, the term of third order in $\vec{e}$.
Before studying our axisymmetric model of the $\delta^{2}$ dynamo in spherical geometry,
we consider a stronger simplified  model in plane geometry,
in order to check whether such a dynamo is in principle possible. For this purpose,
Eq.~(\ref{dif1})
is simplified to
 \begin{equation}
\vec{\mathcal{E}}= -\eta_{0}\nabla\times\langle\vec{B}\rangle+c_{1}\left(\vec{e}\cdot\nabla\right)\langle\vec{B}\rangle+c_{3}\vec{e}\left(\vec{e}\cdot\nabla\right)\left(\vec{e}\cdot\langle\vec{B}\rangle\right)\,,\label{eq:simp-e}
\end{equation}
where $c_1$ and $c_3$ are constants. This MEMF contains an isotropic diffusion term and two terms corresponding
to the $\vec{\Omega}\times\vec{J}$ effect, one of first order and the other
of third order in $\vec{e}$ (their formal structures agree with those of the terms with
coefficients $f_4^{(d)}$ and $f_3^{(d)}$,
respectively, in Eq.~(\ref{dif1})).
The rotation axis coincides with the $z$ axis of a system of Cartesian coordinates $x,y,z$,
and the mean magnetic field is written as
\begin{equation}
 \langle\vec{B}\rangle=\nabla\times\left(a\left(x,z\right)\vec{e}_{y}\right)+b\left(x,z\right)\vec{e}_{y} \,.
\end{equation}
$\vec{e}_{y}$ is the unit vector in the $y$ direction, in which the system is invariant (corresponding
to the $\phi$ direction in the spherical model).
From the mean-field induction equation
\begin{equation}
\frac{\partial\langle\vec{B}\rangle}{\partial t}=\nabla\times\vec{\mathcal{E}}
\end{equation}
we then get
\begin{align}
\left\{ \frac{\partial}{\partial t}-\eta_{0}\left(\frac{\partial}{\partial x^{2}}+\frac{\partial}{\partial z{}^{2}}\right)\right\} a &=  c_{1}\frac{\partial b}{\partial z} \,,\label{eq:simp-d1} \\
\left\{ \frac{\partial}{\partial
    t}-\eta_{0}\left(\frac{\partial}{\partial
      x^{2}}+\frac{\partial}{\partial z{}^{2}}\right)\right\} b 
&=  -c_{1}\frac{\partial^{3}a}{\partial z^{3}}  
-\left(c_{1}+c_{3}\right)\frac{\partial^{3}a}{\partial x^{2}\partial z} \label{eq:simp-d2}\,.
\end{align}
Solutions to Eqs.~(\ref{eq:simp-d1}--\ref{eq:simp-d2}) may be sought in the form
\begin{equation}
 a=\hat{a}\exp\left\{\gamma t+i\left(kx+qz\right)\right\}\,,\quad b=\hat{b}\exp\left\{\gamma t+i\left(kx+qz\right)\right\} \,,
\end{equation}
with $\gamma$ being determined by the condition that the determinant of the resulting system of algebraic
equations vanishes, that is,
\begin{equation}
 \left[\gamma+\eta_{0}\left(k^{2}+q^{2}\right)\right]^{2}+q^{2}\left[c_{1}^{2}\left(k^{2}+q^{2}\right)+c_{1}c_{3}k^{2}\right]=0 \,,
\end{equation}
with roots
\begin{equation}
 \gamma_{1,2}=-\eta_0\left(k^{2}+q^{2}\right) \mp \sqrt{-q^{2}\left[c_{1}^{2}\left(k^{2}+q^{2}\right)+c_{1}c_{3}k^{2}\right]} \,.
\end{equation}
It is seen that  $\gamma$ is real for growing modes, so that only non-oscillatory
dynamo modes can be excited. Furthermore, necessary, but not sufficient conditions
for the existence of a dynamo mode are
(i) $c_1\ne 0$, (ii) $c_3\ne 0$, (iii) $k\ne 0$, (iv) $q\ne 0$ and (v)
\begin{equation}
 -\frac{c_3}{c_1}>\frac{k^2+q^2}{k^2} \,.
\end{equation}
That is, $c_1$ and $c_3$ must have opposite signs and the absolute value of $c_3$ must be larger
than that of $c_1$.
Now in the limit of fast rotation, $\Omega^{*}\gg1$, the signs of $f_{3}^{(d)}$ and $f_{4}^{(d)}$
are opposite,
 namely, for $\Omega^{*}\to\infty$ one gets $f_{3}^{(d)}/f_{4}^{(d)}\to-2$.
Suppose $c_{1}=-c_{3}/2$. Then the (necessary and sufficient) condition for dynamo instability
becomes
\begin{equation}
\frac{c_{3}^2q^2\left(k^{2}-q^{2}\right)}{4\eta_{0}^2\left(k^{2}+q^{2}\right)^2}>1\,.\label{eq:simp-r}
\end{equation}
It can be satisfied if  $\left|k\right|>\left|q\right|$. The
optimal wave number ratio  is $|k|/|q|=\sqrt{3}$, giving dynamo excitation for
 $c_{3}^2>32\eta_{0}^2$. These
findings are largely confirmed by the kinematic $\delta^{2}$ dynamo
model in spherical geometry, for which  we now discuss the eigenmodes of dipolar type.

\begin{figure}
\centering
\includegraphics[width=0.45\columnwidth,angle=0]{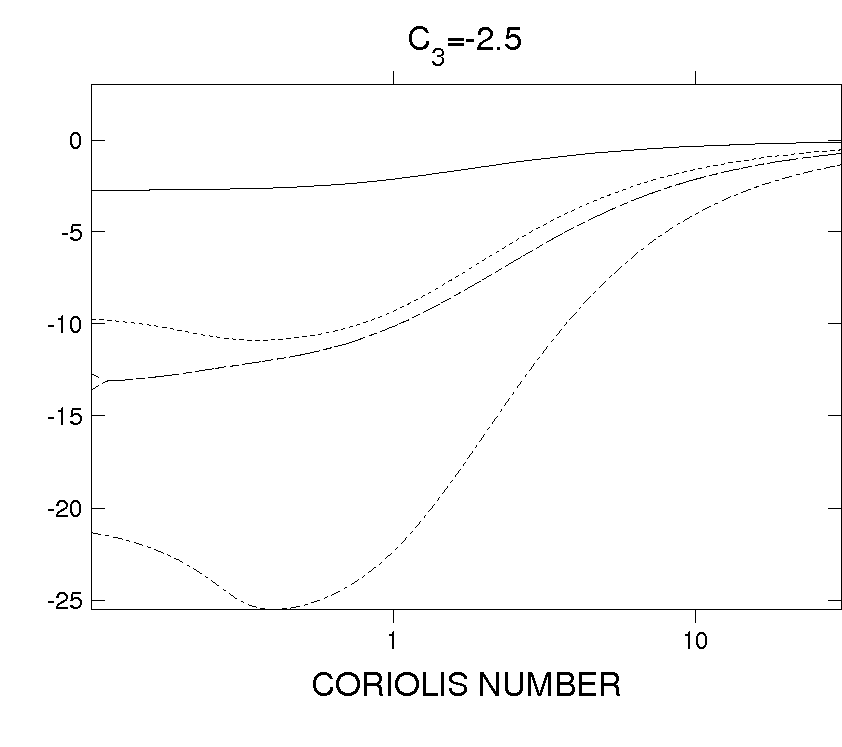}~\includegraphics[width=0.45\columnwidth,angle=0]{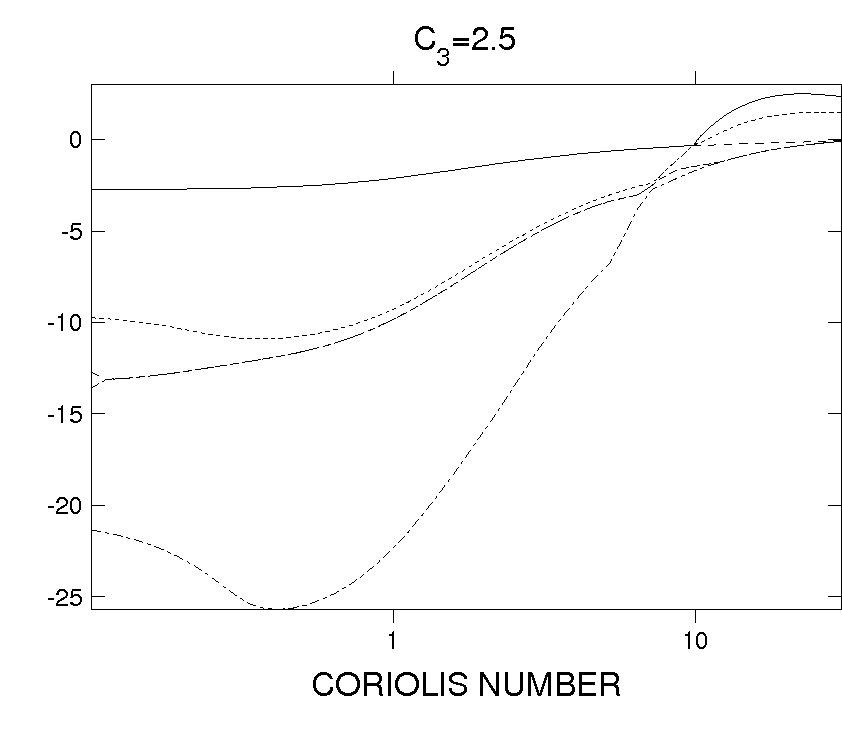}
\caption{\label{fig:stability-D2}Stability diagrams for the $\delta^{2}$ dynamo in spherical geometry
without the effects of anisotropic diffusion and turbulent pumping
for $c_{3}=-2.5$ (left) and  $c_{3}=2.5$ (right). The four largest real parts
of the eigenvalues $\lambda$ are shown
as functions of the Coriolis number. Only dipolar modes are considered.}
\end{figure}

In the spherical model we have left only one of the above two parameters, $c_3$, as a weighting factor for the term
with coefficient $f_3^{(d)}$ (term of third order in $\vec{e}$), see Eqs.~(\ref{eq:er-dlt2}--\ref{eq:et-dlt2}). Besides that we now also vary the Coriolis number
$\Omega^*$.
If we neglect the effects of anisotropic turbulent diffusion (but not those of isotropic turbulent diffusion)
and density stratification (i.e., turbulent pumping)
 in the spherical model, we get steady dynamo modes as  obtained in the
plane model above. Stability diagrams for this case are shown
in Fig.~\ref{fig:stability-D2},
where the left panel shows that
there is no dynamo effect if the coefficients of the terms describing the $\vec{\Omega}\times\vec{J}$ effect
do not have signs in accordance with what
is expected to be necessary for dynamo action from the consideration of the plane model.
Namely, $f_3^{(d)}$ is negative and $f_4^{(d)}$ positive for all $\Omega^*$,
 $\left|f_3^{(d)}/f_4^{(d)}\right|$
monotonically increasing with $\Omega^*$ and $f_3^{(d)}/f_4^{(d)}\approx -2$ for $\Omega^*\gtrsim 5$.
For $c_3=1$ we found
no dynamo instability in the whole $\Omega^{*}$ interval. The
condition for dynamo action was found to be $c_{3}>2$.
Furthermore, a threshold value of $\Omega^*$ must be exceeded.
 The larger
$c_{3}$, the smaller the threshold value of $\Omega^*$.

\begin{figure*}
\includegraphics[angle=0,width=1\textwidth]{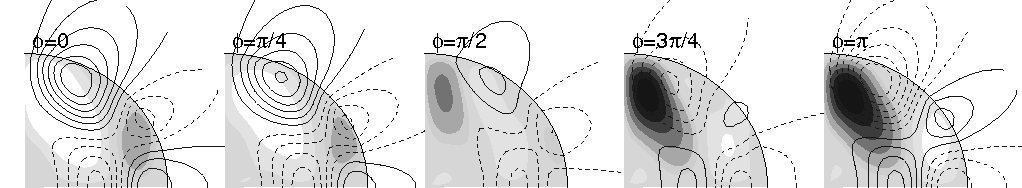}
\caption{\label{fig:Snapshots-d2}Snapshots of the magnetic field in the time-periodic $\delta^2$ dynamo (first unstable mode) over half a cycle for $c_3=2.5$, $\Omega^*=9.1$
in the form of greyscale plots for the toroidal LSMF (bright/dark areas indicate positive/negative
field values) and
field lines of the poloidal LSMF
(solid/dashed lines correspond to clockwise/counter-clockwise field direction).}
\end{figure*}

Including
anisotropic turbulent diffusion decreases the
threshold values of $c_3$ and $\Omega^*$
for instability, while including turbulent transport (pumping),
associated with density stratification, raises them.
Yet, including one of them or both together changes the character of the dynamo bifurcation
from a steady-state to a Hopf bifurcation. Namely,
the two most unstable modes then merge,  yielding a slowly oscillating dynamo mode.
``Slowly'' here means that the time period of the mode is larger than the turbulent
diffusion time of the system. For the parameter choice
$c_{3}=2.5$, $R_\star G=-20$, the threshold value of the Coriolis number is $\Omega^*\approx9.1$, and
the frequency of the obtained dynamo mode is $\omega\approx0.8$ inverse diffusion times.
Fig.~\ref{fig:Snapshots-d2}
shows the evolution of the large-scale toroidal and poloidal field components over half a cycle.
It is seen that
the strongest toroidal and poloidal fields are concentrated close to the
rotation axis in the polar regions (actually the field lines of the poloidal
field are isocontours of the potential $A$, so the strength of the poloidal field is also seen).
The dynamo oscillation closely resembles a standing wave.

\section{Conclusions} \label{sec_conclusions}

We have studied kinematic axisymmetric mean-field dynamo
models in which the $\vec{\Omega}\times\vec{J}$ effect was included as one
of the mechanisms by which the large-scale magnetic field is generated.
The effect was used in an extended form, together with all other effects that result
from the combined action of rotation and a nonuniformity of the large-scale magnetic field,
including anisotropic turbulent diffusion.
Concurrently to the $\alpha$ effect, the $\vec{\Omega}\times\vec{J}$ effect
may, in particular, generate the large-scale poloidal magnetic field of stars.
Our results show that the inclusion of the additional effects alleviates
some of the problems connected with the $\alpha$ effect and $\alpha\Omega$ dynamos.
Compared to the standard $\alpha\Omega$
dynamo, the inclusion of the  $\vec{\Omega}\times\vec{J}$ effect increases the
time period of the dynamo and brings, thus, the models in better agreement
with the solar observations. Furthermore, the large-scale toroidal field comes closer
to the equator, bringing the models in better agreement with the observations
also in this respect. The observed phase relation between the toroidal and poloidal field components 
is correctly reproduced and, in contrast to the $\alpha\Omega$ model, dynamo modes with the
correct (dipolar) parity become unstable first.

We did not explore the possible effects of meridional
circulation in the considered models. Circulation-dominated
dynamo models are very popular presently. They apparently avoid the problems with $\alpha\Omega$
dynamos mentioned above. But they also introduce  (yet) unproven assumptions concerning
the meridional flow
and work only if the turbulent magnetic diffusivity is reduced by about two orders of magnitude
compared to the mixing-length estimates.
We do not claim that the models considered in the present paper are superior
to circulation-dominated dynamo models. Possibly the effects studied here 
and meridional circulation should be combined. We have preliminary results (not published yet)
which indicate that a $\delta\Omega$ dynamo model with meridional
circulation, but without the $\alpha$ effect, admits oscillatory eigenmodes that
give a solar-type dynamo.
Another possible improvement of our model would be the inclusion of the
tachocline.

We also gave a first example of a $\delta^{2}$ dynamo which is solely based
on the joint induction effects of rotation and an inhomogeneity of the large-scale magnetic field,
without differential rotation, the $\alpha$ effect and turbulent pumping. This kind of dynamo may be relevant for fully convective stars. 
We found a dynamo instability both for a strongly simplified
model in plane geometry and for an axisymmetric model for a full sphere. 
The dynamo modes are steady if the effect of anisotropic diffusion is not included. 
In the case of the sphere, the inclusion of anisotropic diffusion
yields a slowly oscillating magnetic field. Including turbulent pumping (due to density stratification,
not part of a pure $\delta^2$ dynamo)
leads to the same result. The period of the oscillation is of the order of the turbulent diffusion
time of the system. For  M dwarfs this is estimated to be  $10\,\mathrm{yr}$ -- $100\,\mathrm{yr}$.
Preliminary results for
combinations of $\alpha^{2}$ and $\delta^{2}$ dynamo models
(not contained in the paper and not yet published elsewhere) indicate that the mixture
of the two effects generally produces oscillatory dynamo modes.
Our aim here was not to provide a dynamo model for fully convective stars, but merely to suggest
the combined effects of rotation and an inhomogeneity of the large-scale magnetic field
as ingredients in future dynamo models.

\begin{acknowledgements}
The work of V.~V.~Pipin was supported by the Russian Foundation for Basis Research
(RFBR) through grants 2258.2008.2 and 07-02-00246.
\end{acknowledgements}

\begin{appendix}
\section*{Appendix}
Here we give the definitions of the functions $f_i^{(a)}$ and $f_i^{(d)}$ that are used in the
representation
of the turbulent electromotive force $\vec{\mathcal{E}}$. For details of the calculations
we refer to \citet{pip08}.

\begin{eqnarray*}
f_{1}^{(a)} & = & \frac{1}
{4\Omega^{*\,2}}\left[\left(\Omega^{*\,2}+3\right)\frac{\arctan\Omega^{*}}
{\Omega^{*}}-3\right]\,,\\
f_{2}^{(a)} & = & \frac{1}
{4\Omega^{*\,2}}\left[\left(\Omega^{*\,2}+1\right)\frac{\arctan\Omega^{*}}
{\Omega^{*}}-1\right]\,,\\
f_{3}^{(a)} & = & \frac{1}
{4\Omega^{*\,2}}\left[\left(\left(\varepsilon-1\right)\Omega^{*\,2}+\varepsilon-3\right)\frac{\arctan\Omega^{*}}
{\Omega^{*}}+3-\varepsilon\right]\,,\\
f_{4}^{(a)} & = & \frac{1}
{6\Omega^{*\,3}}\Bigg[3\left(\Omega^{*4}+6\varepsilon\Omega^{*2}+10\varepsilon-5\right)\frac{\arctan\Omega^{*}}
{\Omega^{*}}\\
&&-\left((8\varepsilon+5)\Omega^{*2}+30\varepsilon-15\right)\Bigg]\,,
\\
f_{5}^{(a)} & = & \frac{1}
{3\Omega^{*\,3}}\left[3\left(\Omega^{*4}+3\varepsilon\Omega^{*2}+5
(\varepsilon-1)\right)\frac{\arctan\Omega^{*}}
{\Omega^{*}}\right.\\
&&-\left((4\varepsilon+5)\Omega^{*2}+15
(\varepsilon-1)\right)\Bigg]\,,\\
f_{6}^{(a)} & = & -\frac{1}
{48\Omega^{*\,3}}\left[3\left(\left(3\varepsilon-11\right)\Omega^{*2}+5\varepsilon-21\right)\frac{\arctan\Omega^{*}}
{\Omega^{*}}\right.\\
&&-\left(4\left(\varepsilon-3\right)\Omega^{*2}+15\varepsilon-63\right)\Bigg]\,,
\\
f_{7}^{(a)} & = & \frac{1}
{48\Omega^{*\,3}}\left[3\left(\left(5\varepsilon+3\right)\Omega^{*2}+11\varepsilon+5\right)\frac{\arctan\Omega^{*}}
{\Omega^{*}}\right.\\
&&-\left(4\left(\varepsilon+1\right)\Omega^{*2}+33\varepsilon+15\right)\Bigg]\,,
\\
f_{8}^{(a)} & = & -\frac{1}
{12\Omega^{*\,3}}\left[3\left(\left(3\varepsilon+1\right)\Omega^{*2}+4\varepsilon-2\right)\frac{\arctan\Omega^{*}}
{\Omega^{*}}\right.\\
&&-\left(5\left(\varepsilon+1\right)\Omega^{*2}+12\varepsilon-6\right)\Bigg]\,,
\\
f_{9}^{(a)} & = & \frac{\varepsilon+1}
{4\Omega^{*}}\left(\frac{\arctan\Omega^{*}}{\Omega^{*}}-1\right)\,,\\
f_{10}^{(a)} & = & \frac{1}
{3\Omega^{*\,3}}\left[3\left(\Omega^{*2}+1\right)\left(\Omega^{*2}+\varepsilon-1\right)\frac{\arctan\Omega^{*}}
{\Omega^{*}}\right.\\
&&-\left(\left(2\varepsilon+1\right)\Omega^{*2}+3\varepsilon-3\right)\Bigg]\,,
\\
f_{11}^{(a)} & = & \frac{1}
{6\Omega^{*\,3}}\left[3\left(\Omega^{*2}+1\right)\left(\Omega^{*2}+2\varepsilon-1\right)\frac{\arctan\Omega^{*}}
{\Omega^{*}}\right.\\
&&-\left(\left(4\varepsilon+1\right)\Omega^{*2}+6\varepsilon-3\right)\Bigg]\,.
\end{eqnarray*}
\begin{eqnarray*}
f_{1}^{(d)} & = & \frac{1}
{2\Omega^{*\,3}}\Bigg[\left(\varepsilon+1\right)\Omega^{*\,2}+3\varepsilon\\
&&-\left(\left(2\varepsilon+1\right)\Omega^{*\,2}+3\varepsilon\right)\frac{\arctan\left(\Omega^{*}\right)}
{\Omega^{*}}\Bigg]\,,\\
f_{2}^{(d)} & = & \frac{1}
{4\Omega^{*\,2}}\left[\left(\left(\varepsilon-1\right)\Omega^{*\,2}+3\varepsilon+1\right)\frac{\arctan\left(\Omega^{*}\right)}
{\Omega^{*}}\right.\\
&&-\left(3\varepsilon+1\right)\Bigg]\,,\\
f_{3}^{(d)} & = & \frac{1}
{2\Omega^{*\,3}}\left[3\left(3\Omega^{*\,2}+5\right)\frac{\arctan\left(\Omega^{*}\right)}
{\Omega^{*}}-\left(4\Omega^{*\,2}+15\right)\right]\,,\\
f_{4}^{(d)} & = & \frac{1}
{2\Omega^{*\,3}}\left[\left(2\Omega^{*\,2}+3\right)-3\left(\Omega^{*\,2}+1\right)\frac{\arctan\left(\Omega^{*}\right)}
{\Omega^{*}}\right]\,.\end{eqnarray*}

\end{appendix}


\begin{thebibliography}{47}
\expandafter\ifx\csname natexlab\endcsname\relax\def\natexlab#1{#1}\fi

\bibitem[{Boyd(2001)}]{boy01}
Boyd, J.~P. 2001, {C}hebyshev and {F}ourier Spectral Methods, 2nd edn.
  (Mineola, N. Y.: Dover Publications)

\bibitem[{Brandenburg(2005)}]{bra05}
Brandenburg, A. 2005, \apj, 625, 539

\bibitem[{Brandenburg \& Subramanian(2005)}]{brasub05}
Brandenburg, A. \& Subramanian, K. 2005, \physrep, 417, 1

\bibitem[{Browning(2008)}]{bro08}
Browning, M.~K. 2008, \apj, 676, 1262

\bibitem[{Chabrier \& K\"uker(2006)}]{chakuk06}
Chabrier, G. \& K\"uker, M. 2006, \aap, 446, 1027

\bibitem[{Dikpati \& Gilman(2007)}]{dikgil07}
Dikpati, M. \& Gilman, P.~A. 2007, New J. Phys., 9, 297

\bibitem[{Dikpati {et~al.}(2005)Dikpati, Rempel, Gilman, \&
  Mac{G}regor}]{diketal05}
Dikpati, M., Rempel, M., Gilman, P.~A., \& Mac{G}regor, K.~B. 2005, \aap,
  437, 699

\bibitem[{Dobler {et~al.}(2006)Dobler, Stix, \& Brandenburg}]{dobstibra06}
Dobler, W., Stix, M., \& Brandenburg, A. 2006, \apj, 638, 336

\bibitem[{{Eggleton}(1971)}]{egg71}
{Eggleton}, P.~P. 1971, \mnras, 151, 351

\bibitem[{{Eggleton} \& {Kiseleva-Eggleton}(2002)}]{eggkis02}
{Eggleton}, P.~P. \& {Kiseleva-Eggleton}, L. 2002, \apj, 575, 461

\bibitem[{Elstner \& R\"udiger(2007)}]{elsrud07}
Elstner, D. \& R\"udiger, G. 2007, Astron. Nachr., 328, 1130

\bibitem[{{Galloway} \& {Weiss}(1981)}]{galwei81}
{Galloway}, D.~J. \& {Weiss}, N.~O. 1981, \apj, 243, 945

\bibitem[{{Godier} \& {Rozelot}(2000)}]{godroz00}
{Godier}, S. \& {Rozelot}, J.-P. 2000, \aap, 355, 365

\bibitem[{Guerrero \& {de Gouveia Dal Pino}(2008)}]{guegou08}
Guerrero, G. \& {de Gouveia Dal Pino}, E.~M. 2008, \aap, 485, 267

\bibitem[{Izzard \& Glebbeek(2006)}]{izzgle06}
Izzard, R.~G. \& Glebbeek, E. 2006, \na, 12, 161

\bibitem[{Jouve {et~al.}(2008)Jouve, Brun, Arlt, Brandenburg, Dikpati, Bonanno,
  K\"apyl\"a, Moss, Rempel, Gilman, Korpi, \& Kosovichev}]{jouetal08}
Jouve, L., Brun, A.~S., Arlt, R., {et~al.} 2008, \aap, 483, 949

\bibitem[{{Kichatinov} \& {Pipin}(1993)}]{kitpip93}
{Kichatinov}, L.~L. \& {Pipin}, V.~V. 1993, \aap, 274, 647

\bibitem[{Kitchatinov(2002)}]{kit02}
Kitchatinov, L.~L. 2002, \aap, 394, 1135

\bibitem[{Kitchatinov(2004)}]{kit04}
Kitchatinov, L.~L. 2004, in {NATO} Science Series II: Mathematics, Physics and
  Chemistry, Vol. 124, Turbulence, Waves and Instabilities in the Solar Plasma,
  ed. R.~Erd\'elyi, K.~Petrovay, B.~Roberts, \& M.~J. Aschwanden (Dordrecht:
  Kluwer), 81--96

\bibitem[{Kitchatinov {et~al.}(1994)Kitchatinov, Pipin, \&
  R\"udiger}]{kitpiprud94}
Kitchatinov, L.~L., Pipin, V.~V., \& R\"udiger, G. 1994, Astron. Nachr., 315,
  157

\bibitem[{K\"ohler(1973)}]{koh73}
K\"ohler, H. 1973, \aap, 25, 467

\bibitem[{Krause \& R\"adler(1980)}]{krarad80}
Krause, F. \& R\"adler, K.-H. 1980, Mean-Field Magnetohydrodynamics and Dynamo
  Theory (Berlin: Akademie-Verlag)

\bibitem[{{K{\"u}ker} \& {R{\"u}diger}(1999)}]{kukrud99}
{K{\"u}ker}, M. \& {R{\"u}diger}, G. 1999, \aap, 346, 922

\bibitem[{Kuzanyan {et~al.}(2006)Kuzanyan, Pipin, \& Seehafer}]{kuzpipsee06}
Kuzanyan, K.~M., Pipin, V.~V., \& Seehafer, N. 2006, \solphys, 233, 185

\bibitem[{Ossendrijver(2003)}]{oss03}
Ossendrijver, M. 2003, \aapr, 11, 287

\bibitem[{Parker(1955)}]{par55}
Parker, E.~N. 1955, \apj, 122, 293

\bibitem[{Parker(1979)}]{par79}
Parker, E.~N. 1979, Cosmical Magnetic Fields (Oxford: Clarendon Press)

\bibitem[{Pipin(2007)}]{pip07}
Pipin, V.~V. 2007, Astron. Rep., 51, 411

\bibitem[{Pipin(2008)}]{pip08}
Pipin, V.~V. 2008, Geophys. Astrophys. Fluid Dynam., 102, 21

\bibitem[{R\"adler(1969)}]{rad69}
R\"adler, K.-H. 1969, Monatsber. Dtsch. Akad. Wiss. Berlin, 11, 194, in German,
  English translation in P. H. Roberts and M. Stix, Report No. NCAR-TN/IA-60
  (1971)

\bibitem[{R\"adler(1980)}]{rad80}
R\"adler, K.-H. 1980, Astron. Nachr., 301, 101

\bibitem[{R\"adler(2000)}]{rad00}
R\"adler, K.-H. 2000, in Lecture Notes in Physics, Vol. 556, From the Sun to
  the Great Attractor: 1999 Guanajuato Lectures on Astrophysics, ed. D.~Page \&
  J.~G. Hirsch (New York: Springer), 101 -- 172

\bibitem[{R\"adler {et~al.}(2003)R\"adler, Kleeorin, \&
  Rogachevskii}]{radklerog03}
R\"adler, K.-H., Kleeorin, N., \& Rogachevskii, I. 2003, Geophys. Astrophys.
  Fluid Dynam., 97, 249

\bibitem[{R\"adler \& Rheinhardt(2007)}]{radrhe07}
R\"adler, K.-H. \& Rheinhardt, M. 2007, Geophys. Astrophys. Fluid Dynam., 101,
  117

\bibitem[{R\"adler \& Stepanov(2006)}]{radste06}
R\"adler, K.-H. \& Stepanov, R. 2006, \pre, 73, 056311

\bibitem[{Rogachevskii \& Kleeorin(2003)}]{rogkle03}
Rogachevskii, I. \& Kleeorin, N. 2003, \pre, 68, 036301

\bibitem[{Rogachevskii \& Kleeorin(2004)}]{rogkle04}
Rogachevskii, I. \& Kleeorin, N. 2004, \pre, 70, 046310

\bibitem[{R\"udiger \& Hollerbach(2004)}]{rudhol04}
R\"udiger, G. \& Hollerbach, R. 2004, The Magnetic Universe (Weinheim:
  Wiley-VCH)

\bibitem[{Schou {et~al.}(1998)Schou, Antia, Basu, Bogart, Bush, Chitre,
  Christensen-Dalsgaard, {Di Mauro}, Dziembowski, Eff-Darwich, Gough, Haber,
  Hoeksema, Howe, Korzennik, Kosovichev, Larsen, Pijpers, Scherrer, Sekii,
  Tarbell, Title, Thompson, \& Toomre}]{schouetal98}
Schou, J., Antia, H.~M., Basu, S., {et~al.} 1998, \apj, 505, 390

\bibitem[{Sch\"ussler(1980)}]{schus80}
Sch\"ussler, M. 1980, \nat, 288, 150

\bibitem[{Sch\"ussler \& Ferriz-Mas(2003)}]{schusfer03}
Sch\"ussler, M. \& Ferriz-Mas, A. 2003, in Advances in Nonlinear Dynamos, ed.
  A.~Ferriz-Mas \& M.~N{\'u}{\~n}ez (London: Taylor \& Francis), 123--146

\bibitem[{Seehafer {et~al.}(2003)Seehafer, Gellert, Kuzanyan, \&
  Pipin}]{seeetal03}
Seehafer, N., Gellert, M., Kuzanyan, K.~M., \& Pipin, V.~V. 2003, Adv. Space Res., 32, 1819

\bibitem[{{Spiegel} \& {Weiss}(1980)}]{spiwei80}
{Spiegel}, E.~A. \& {Weiss}, N.~O. 1980, \nat, 287, 616

\bibitem[{Steenbeck {et~al.}(1966)Steenbeck, Krause, \& R\"adler}]{stk66}
Steenbeck, M., Krause, F., \& R\"adler, K.-H. 1966, Z. Naturforsch., 21a, 369

\bibitem[{Stix(1976)}]{sti76}
Stix, M. 1976, \aap, 47, 243

\bibitem[{Stix(2002)}]{sti02}
Stix, M. 2002, The Sun. An Introduction, 2nd edn. (Berlin: Springer)

\bibitem[{Yoshimura(1975)}]{yos75}
Yoshimura, H. 1975, \apj, 201, 740

\end{thebibliography}

\end{document}